\documentclass[12pt,a4paper]{article}

\usepackage{latexsym}
\usepackage{amsmath}
\usepackage{amsfonts}
\usepackage{amssymb}

\newcommand{\be}{\begin{equation}}
\newcommand{\ee}{\end{equation}}
\newcommand{\bea}{\begin{eqnarray}}
\newcommand{\eea}{\end{eqnarray}}

\def\1{{\mbox{\boldmath $1$}}}          %
\def\0{{\mbox{\boldmath $0$}}}          %
\begin{document}

\begin{center}
{\Large \bf  
E pur si muove!}
\end{center}

\vspace{0.2cm}

\begin{center}
 C. Klim\v c\'\i k${}^b$  \\

\bigskip

${}^b$ Institute de math\'ematiques de Luminy,
 \\ 163, Avenue de Luminy, \\ 13288 Marseille, France\\
 e-mail: klimcik@iml.univ-mrs.fr

\bigskip

\end{center}

\vspace{0.2cm}

\begin{abstract} We comment a debate  on quantum physics  in Slovak TV in which several physicists advocated a point of view that   "an electron does not move  around the nucleus".
 
\end{abstract}

 \vskip1cm

This note  records some musings about the quantum world which could perhaps help
beginners  in making first steps in quantum mechanics. It is principally addressed to students of  Bratislava University but it may  be of some interest more generally, e.g. to people fascinated by real or apparent paradoxes of  the quantum world.
In particular, on the example below, I  point out   disagreements between me and some teachers from Bratislava University about the interpretation of  some  basic concepts  of quantum mechanics

In what follows, I shall summarise and comment a part of  \v Stefan Hr\'\i b's  TV show "The evening under the lamp"   which is mainly devoted to topical  political  issues in Slovakia, but    that particular evening it was  devoted to  "The smallest parts of the matter" \cite{YT}. Among invited guests there was a teacher
of quantum field theory at Bratislava University  Martin Moj\v zi\v s, two authors of a textbook on quantum mechanics Vladim\'\i r  \v Cern\'y and J\'an Pi\v s\'ut
and an experimental physicist Juraj Bracin\'\i k. Among other topics, it was discussed a question whether an electron moves around the nucleus.  At some point Martin Moj\v zi\v s  said:

{\it " ... when the electron is in the ground state then it is in it  and it is all the time in this state forever. If the atom did not interact with anything else it is in this state forever".}

 The TV host  \v Stefan Hr\'\i b then asked a truly  outstanding  question for a layman, in which he has shown his remarkable
talent, intelligence and intuition. Indeed,  in accordance with the very principles of the quantum theory (at least as I understand them), he 
half-stated half-asked:

{\it "This means that the electron moves in the same way forever?"}

To my big surprise Martin Moj\v zi\v s has disagreed with this  interpretation and he objected:

{\it  " Moves" is a wrong word, in this case misleading.
It is  forever in the same state. "}

When   \v Stefan  Hr\'\i b's  opened his mouth to protest Martin Moj\v zi\v s  did not let him speak and   insisted:

{\it "Moves" is  a  misleading word. OK,  it can be used but it obscures much more than it elucidates.}

Then all remaining  physicists have  supported the point of view of Martin Moj\v zi\v s,  inspite of the fact that  \v Stefan Hr\'\i b, visibly confused, was fighting hard.  At the very end    the  TV host  said reluctantly :

{\it  "So we have abandoned the idea that the electron is moving around the nucleus... "}

Then he made  a tiny  break as if he expected a  late sign of disapproval  from  the present physicists against such a "blasphemy".  However, the experts did not  react so \v Stefan Hr\'\i b continued: 

{\it "Well, it is quite sad to tell the truth..."}

 Again an infinitesimal   break but the scientists  remained inflexible.  The TV host, visibly disappointed, then gave up: 

{\it "Ok then..."}
 
 \vskip 1pc

 Honestly, at the very beginning I did not understand why the colleagues from Bratislava were advocating such a strange point of view.
 Then it came into my mind that, perhaps,   they do not understand the fact that if the electron remains all the time in the same state it does not mean at all  that it  is not moving!  
 
 To explain it, consider a free particle on a line or, better, on  a large circle to avoid problems with normalisation of the wave function.
 Take the  wave function of the particle   proportional to $e^{ip_1x}$ for some non-zero momentum $p_1$  so the state described by this wave function REMAINS THE SAME all the time.  However,  it is obvious that  the state $e^{ip_1x}$ describes the MOVING  particle  since it is in the eigenstate of the momentum operator!  The solution of this "paradox" is precisely that proposed by the  TV host  \v Stefan Hr\'\i b: the state of the free particle does not change with time since the particle  moves all the time in the same way!
 
 A similar analysis applies for a rotational motion. Although Martin Moj\v zi\v s was speaking about the  ground  state of the hydrogen atom (presumably with "spinless" electron),  where there is no angular motion, the other physicists did have a possibility to clarify things by saying    something like: {\it   "The  operator of angular momentum commutes with the Coulomb
 Hamiltonian therefore there are plenty excited angular momentum eigenstates  which do not evolve with time but still they clearly describe the rotational motion of the electron around the nucleus."}
 
 After all, even in the ground state the electron does move albeit in the radial way. To see that, it is sufficient to argue that the mean value of the
 kinetic energy in the ground state does not vanish. This can be seen without any calculation since the normalizability of the bounded
 ground state makes possible the integration "per partes" and the mean value of the kinetic energy $<T>$  is thus proportional to
 
$$<T> \sim \int \nabla \psi^*\nabla \psi.$$
Obviously, this  is a strictly positive quantity for  the (nonconstant)  ground state wave function $\psi$.

I think that the things can be made even clearer if we consider the classical limit of the quantum mechanical concept of the "state".
Indeed, following the folkloric knowledge (nevertheless not mentioned in the textbook of Vladim\'\i r \v Cern\'y and J\'an Pi\v s\'ut),  the classical
limit of the quantum state is a maximally isotropic  submanifold   of the classical phase space\footnote{Maximally isotropic submanifold of  a $2n$-dimensional phase space is an $n$-dimensional submanifold on which the symplectic form vanishes.}.  If we view
the classical state as a point of the phase space, we observe that the classical limit of the quantum state is a \underline{collection}  of the classical states.
If we consider classical  Hamiltonian equations  of motion, it may very well happen, that  solutions with initial values   on the maximally isotropic submanifold will remain contained in  this submanifold  for all subsequent times.   Such  maximally isotropic submanifold may be then called "stationary" and, upon quantization, it gives rise to a stationary quantum state which does not evolve with time.  However, the classical motion within the stationary maximally isotropic submanifold is still present and it is  present also in the corresponding quantum stationary state in the subtle way mentioned  above.

\noindent {\bf Example:}

Consider a classical  free particle in one dimension. The coordinates of the phase space are $p$ and $q$ and the symplectic form is
$dp\wedge dq$. Consider a constant number $p_1$. On the one-dimensional  submanifold defined by the relation $p=p_1$ the symplectic form vanishes therefore this submanifold is maximally isotropic. It is also stationary since the solution of the free equations (with the Hamiltonian $H=p^2/2$)  reads
$p(t)=p_1$, $q(t)=q_1+p_1t$ therefore it respects all the times the constraint $p=p_1$.  The quantization of the stationary maximally isotropic submanifold $p=p_1$ is the wave function $e^{ip_1x}$ mentioned above and it encodes  in it the free motion $q(t)=q_1+p_1t$
in a subtle way, that is,  $e^{ip_1x}$ is the eigenstate of the momentum operator with the eigenvalue $p_1$.

\end{document}